\documentclass[sigconf]{acmart} 
\AtBeginDocument{%
  }

\copyrightyear{2026}
\acmYear{2026}
\setcopyright{cc}
\setcctype{by-nc-nd}
\acmConference[CHI EA '26]{Extended Abstracts of the 2026 CHI Conference on Human Factors in Computing Systems}{April 13--17, 2026}{Barcelona, Spain}
\acmBooktitle{Extended Abstracts of the 2026 CHI Conference on Human Factors in Computing Systems (CHI EA '26), April 13--17, 2026, Barcelona, Spain}
\acmDOI{10.1145/3772363.3798725}
\acmISBN{979-8-4007-2281-3/2026/04}

\usepackage{subcaption}

\begin{document}

\title{GazeSync: A Mobile Eye-Tracking Tool for Analyzing Visual Attention on Dynamically Manipulated Content}

\author{Yaxiong Lei}
\email{yl212@st-andrews.ac.uk}
\orcid{0000-0002-0697-7942}
\affiliation{%
  \institution{University of St Andrews}
  \city{St Andrews}
  \country{UK}
}
\affiliation{%
  \institution{University of Essex}
  \city{Colchester}
  \country{UK}
}

\author{Rishab Talwar}
\email{rt81@st-andrews.ac.uk}
\orcid{0009-0006-7998-8964}
\affiliation{%
  \institution{University of St Andrews}
  \city{St Andrews}
  \country{UK}
}

\author{Shijing He}
\email{shijing.he@kcl.ac.uk}
\orcid{0000-0003-3697-0706}
\affiliation{%
 \institution{King's College London}
 \city{London}
 \country{UK}
}

\author{Xinya Gong}
\email{xg31@st-andrews.ac.uk}
\orcid{0009-0005-6414-9351}
\affiliation{%
  \institution{University of St Andrews}
  \city{St Andrews}
  \country{UK}
}

\author{Yuheng Wang}
\email{yw99@st-andrews.ac.uk}
\orcid{0000-0003-3335-8706}
\affiliation{%
  \institution{University of St Andrews}
  \city{St Andrews}
  \country{UK}
}

\author{Xudong Cai}
\email{xudong.cai@kcl.ac.uk}
\orcid{0009-0002-4376-5022}
\affiliation{%
 \institution{King's College London}
 \city{London}
 \country{UK}
}

\author{Zhongliang Guo}
\email{zg34@st-andrews.ac.uk}
\orcid{0000-0002-6025-3021}
\affiliation{%
  \institution{University of St Andrews}
  \city{St Andrews}
  \country{UK}
}

\author{Juan Ye}
\email{Juan.Ye@st-andrews.ac.uk}
\orcid{0000-0002-2838-6836}
\affiliation{%
 \institution{University of St Andrews}
  \city{St Andrews}
  \country{UK}
}

\renewcommand{\shortauthors}{Yaxiong Lei et al.}

\begin{abstract}

Conventional mobile eye-tracking maps gaze to static screen coordinates, failing to capture user attention when content is dynamic. As users pinch, zoom, and rotate images, static coordinates lose their semantic meaning relative to the underlying visual content. To address this methodological gap, we present \textit{GazeSync}, a reusable mobile system that synchronizes on-device gaze estimation with real-time image transformation matrices (scale, rotation, and translation). By logging gaze coordinates alongside precise UI states, GazeSync enables the accurate reconstruction of \textit{image-relative} attention patterns, decoupling visual attention from device interaction. We validate our end-to-end toolchain through a formative study involving guided manipulation, reading, and visual search tasks. Our results demonstrate GazeSync's ability to recover ground-truth gaze locations on transforming content, explicitly showing how it outperforms static baselines, while also surfacing critical boundaries regarding calibration drift and reconstruction fragility under compound manipulations.
\end{abstract}




\begin{teaserfigure}
  \includegraphics[width=\textwidth]{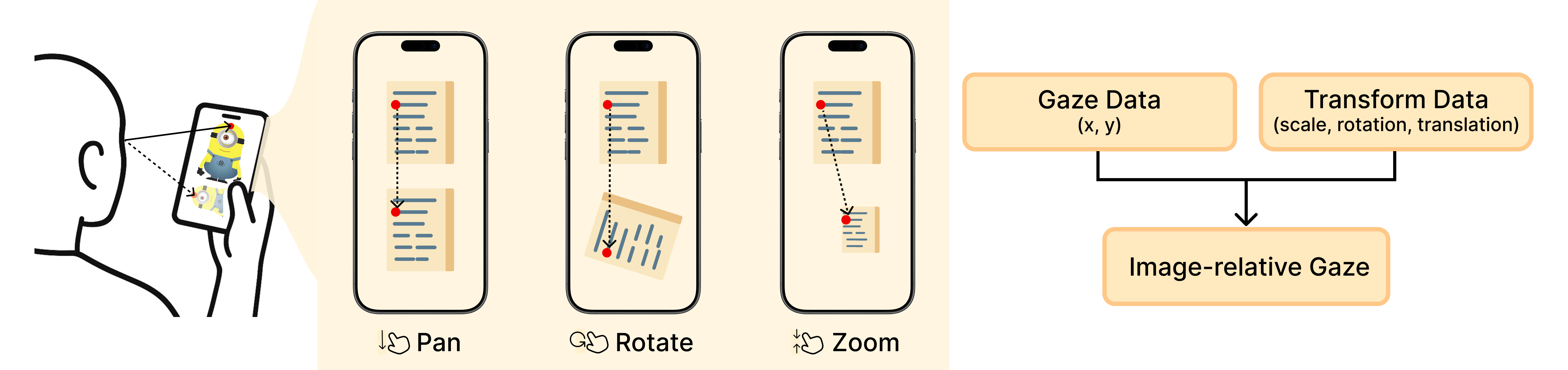}
  \caption{Overview of GazeSync. The system synchronizes on-device gaze estimates with image transformation parameters (pan, rotation, and zoom) to reconstruct image-relative gaze, enabling attention analysis in content coordinates during dynamic manipulation.}
  \Description{A teaser illustration showing a user viewing a smartphone while interacting with an image. Three phone screens depict pan, rotate, and zoom actions. A diagram on the right shows two inputs—gaze data (x, y) and transform data (scale, rotation, translation)—combined to produce image-relative gaze.}
  \label{fig:teaser}
\end{teaserfigure}

\maketitle

\section{Introduction}
Understanding visual attention is fundamental to Human-Computer Interaction~\cite{lei2023end}, with applications in usability testing~\cite{russ2011modeling, diaz2023continuous}, education~\cite{halszka2017eye}, privacy protection~\cite{lei2023protecting, he2025identity, katsini2020role} and medical/clinical workflows~\cite{tahri2023eye}. With smartphones becoming the dominant computing platform, mobile eye-tracking has attracted sustained interest~\cite{bulling2010toward, namnakani2023gaze}. However, \emph{mobile viewing is rarely static}: users continuously pinch-to-zoom, pan, and rotate content (maps, photos, PDFs, scans). This creates a core methodological gap. Most mobile gaze estimation pipelines—ranging from early prototypes~\cite{mercier2024quantifying} to modern deep models~\cite{gunawardena2024deep, lei2023end, cheng2024appearance}—report gaze in \emph{screen coordinates}. Once content is manipulated, the same screen point no longer corresponds to the same underlying content feature. Consequently, screen-based gaze logs can become misleading for studying visual behaviour during active exploration.

\textbf{Contributions.} We present \textbf{GazeSync}, an end-to-end mobile toolchain for \emph{content-relative} attention analysis on dynamically manipulated images. The core idea is simple but missing in many mobile gaze pipelines: we \emph{synchronise each gaze sample with the exact image transformation state} (scale, rotation, translation) and reconstruct gaze in the original image coordinate system. This allows researchers to answer the methodological question: \emph{``Where on the underlying image is the user attending while zooming, panning, and rotating?''} Concretely, GazeSync contributes: (i) a unified logging schema that pairs gaze with transformation state, (ii) an inverse-transform reconstruction pipeline for image-relative gaze, and (iii) lightweight in-app analysis for replay, overlays, and heatmaps. We further report a formative technical validation that illustrates both stable operating cases and practical failure modes (calibration drift and compound-transform fragility), to clarify current usage boundaries and future improvement directions.

\section{Related Work}
Traditional eye tracking relies on specialised, high-precision lab hardware and controlled setups, e.g., pupil centre-corneal reflection-based systems~\cite{duchowski2017eye, guestrin2006general, gonzalez2023eye}. These trackers provide strong spatial/temporal accuracy, but are costly, require careful setup, and do not naturally support in-the-wild mobile use~\cite{valliappan2020accelerating}. Gaze estimation on commodity devices has advanced rapidly, producing a large literature and several surveys (see, e.g.,~\cite{lei2023end, cheng2024appearance}). Broadly, methods fall into two paradigms: \textit{model-based (geometric)} approaches, which infer gaze using 3D eye/corneal models and camera-head geometry~\cite{strupczewski2016geometric, zhou20163d}, and \textit{appearance-based (learning)} approaches, which learn mappings from face/eye imagery to gaze direction or point-of-gaze~\cite{cheng2024appearance, zhang2015appearance, sun2017deep}. On mobile devices, milestones such as iTracker~\cite{krafka2016eye} and large-scale efforts like GoogleGaze~\cite{valliappan2020accelerating} have greatly improved accessibility. However, most mobile pipelines still report gaze in \emph{screen coordinates} and implicitly assume a \emph{static screen plane} for downstream analysis, which becomes brittle once users actively pan, zoom, and rotate content. A substantial HCI literature treats gaze as an input modality to support interaction~\cite{lei2023end}, including gaze-directed zooming~\cite{stellmach2012investigating}, foveated rendering~\cite{pai2016gazesim}, and gaze-based control techniques such as dwell, smooth pursuit, and gaze gestures~\cite{lei2023dynamicread,ramirez2021gazehold, lei2026people}. These systems typically aim to improve interaction efficiency or reduce manual effort, and they often operate in the device/ screen coordinate frame. In contrast to gaze-as-input, we target a methodological need for \emph{observing and analysing natural visual behaviour} while users manipulate visual content via touch. When the viewed content is continuously transformed, screen-coordinate gaze logs conflate attention with interaction state: the same screen location can correspond to different underlying content regions. GazeSync addresses this gap by synchronising each gaze sample with the corresponding transformation parameters (scale/ rotation/ translation) and reconstructing \emph{image-relative} gaze via inverse transforms, enabling content-relative heatmaps, trace overlays, and replay that remain meaningful during pan/zoom/rotate exploration.

\section{The GazeSync System}

GazeSync is a Flutter-based iOS application with a lightweight local logging server (Python/FastAPI). The design goals are: \textit{(i)} robust synchronisation of gaze and interaction state, and \textit{(ii)} rapid validation via in-app analysis. The gaze tracking module uses a commercial on-device gaze SDK (Eydid/SeeSo~\cite{eyedidsdk}) to estimate gaze in real time without storing or transmitting camera imagery. The app includes a short five-point calibration procedure. A live gaze-dot preview supports calibration quality checking, and recalibration is available at any time during the session.


\begin{figure*}[t]
  \centering
  \includegraphics[width=0.95\textwidth]{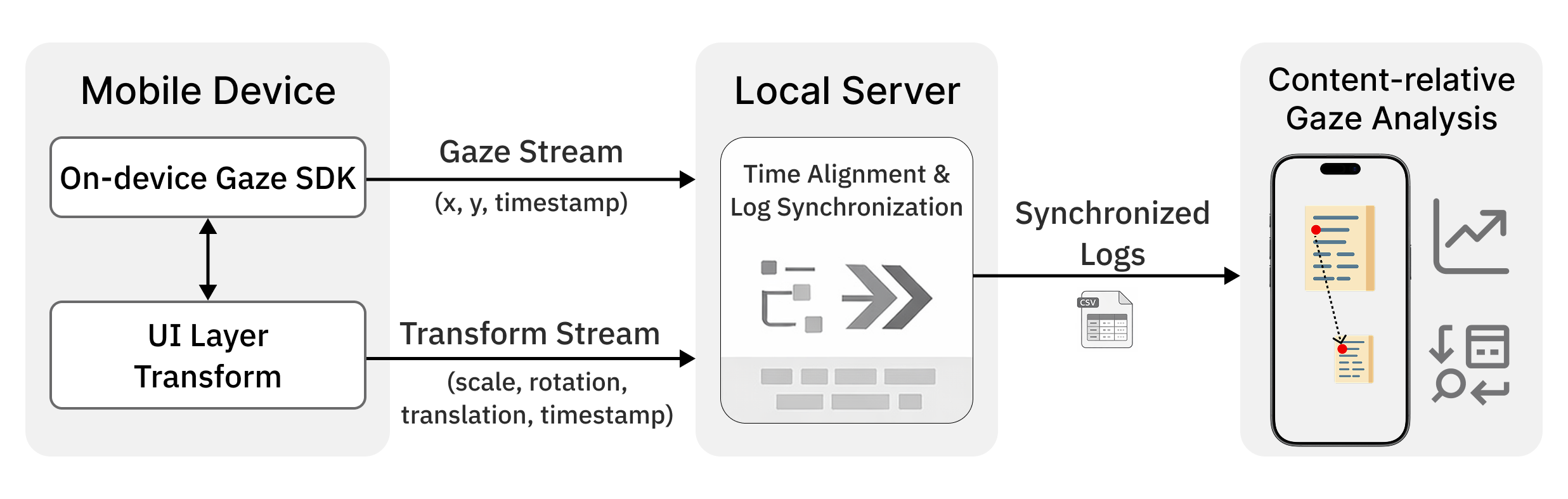}
  \caption{GazeSync architecture. On-device gaze estimation produces screen-coordinate gaze samples, while the Flutter UI logs image transformation parameters. The two asynchronous streams are time-aligned and merged by a local server into synchronized records for replay and content-relative analysis.}
  \Description{A system architecture diagram with three blocks arranged left to right. The first block is the mobile device, containing an on-device gaze SDK and a UI transform layer. Two labelled streams leave the device: a gaze stream with x, y, and timestamp, and a transform stream with scale, rotation, translation, and timestamp. Both enter a local server block that performs time alignment and log synchronization. A synchronized log output is then sent to a content-relative gaze analysis block, illustrated with a phone screen and analysis icons.}
  \label{fig:architecture}
\end{figure*}

\subsection{Core Contribution: Unified, Synchronised Logging}
The core technical contribution is the \textit{unified log row} that pairs gaze and content transformation state. During tasks, the system records two streams simultaneously. First, the \textit{Gaze stream} captures normalised screen gaze coordinates $(x, y)$ with timestamps. Second, the \textit{Transform stream} logs image manipulation state (scale $s$, rotation $\theta$, translation $(t_x, t_y)$). A local server merges these into a single \texttt{combined} record keyed by participant and task, producing rows of the form $(\texttt{pid}, \texttt{task}, x, y, t, s, \theta, t_x, t_y)$, where $t$ is the Unix timestamp (ms) of the gaze sample.

\textit{Synchronisation and merge policy.}
In our implementation, each gaze event triggers a paired read of the current transform state within the same UI callback, and both are sent to a local logging server. The server merges the two messages into a single \texttt{combined} row keyed by participant and task. Concretely, the server temporarily buffers the most recent gaze event and merges it when the corresponding transform message arrives (and vice versa). To make the merge robust to network jitter, we enforce a timestamp tolerance $\Delta = 50$\,ms: a gaze sample at time $t$ is paired with the nearest transform state within $|t-t'|\le\Delta$. If no match exists, the sample is discarded and reported in a log-quality summary. In our formative validation, 96.5\% of samples successfully synchronized within the 50 ms tolerance, with only 3.5\% of samples discarded, demonstrating the reliability of the event-based merge policy under typical mobile usage.

\subsection{Image-Relative Gaze Reconstruction}
Screen gaze is insufficient when content moves under the finger. GazeSync reconstructs image-relative gaze by applying the inverse of the logged transform matrix.

\subsubsection{Coordinate Systems Mapping}
We distinguish three 2D coordinate frames and map between them as follows:

\textbf{(1) Screen Coordinates.} The gaze estimator outputs normalised coordinates $(x_n, y_n)$. We convert these to absolute screen pixels $(W, H)$:
\[
\mathbf{p}_s = [x_n W,\ y_n H,\ 1]^\top
\]

\textbf{(2) Viewport Coordinates.} The image is rendered inside a widget with top-left origin $(o_x, o_y)$. We first map the screen gaze to the widget's local frame:
\[
\mathbf{p}_v = [x_s - o_x,\ y_s - o_y,\ 1]^\top
\]

\textbf{(3) Intrinsic Image Coordinates.} Finally, we map the local viewport coordinates to the intrinsic resolution of the source image $(w_i, h_i)$ based on the displayed dimensions $(w_d, h_d)$:
\[
x_i = x_v \cdot \frac{w_i}{w_d}, \quad y_i = y_v \cdot \frac{h_i}{h_d}
\]
The resulting $(x_i, y_i)$ represents the pixel on the original full-resolution image that corresponds to the user's gaze.

\subsubsection{Transformation Logic}
Let $\mathbf{p}_s$ be the gaze point in homogeneous screen coordinates. To find the point $\mathbf{p}_i$ on the underlying image, we apply the inverse of the active transformation matrix $\mathbf{M}$:
\[
\mathbf{p}_i = \mathbf{M}^{-1}\mathbf{p}_s
\]
We model the UI transformation $\mathbf{M}$ as a composition of user translation $\mathbf{T}$, rotation $\mathbf{R}$, and scaling $\mathbf{S}$. Crucially, rotation and scaling occur around the image center (pivot). Therefore, the composite matrix is formed by:
\[
\mathbf{M} = \mathbf{T}(t_x,t_y) \cdot \mathbf{T}_{center} \cdot \mathbf{R}(\theta) \cdot \mathbf{S}(s) \cdot \mathbf{T}^{-1}_{center}
\]
where $\mathbf{T}_{center}$ represents the translation to the visual center of the image widget. This formulation ensures that coordinate reconstruction accounts for the correct anchor point of the user's gestures.

\subsection{Integrated In-App Analysis for Rapid Validation}
GazeSync includes an \textbf{Analysis Mode} that loads logged sessions and provides several visualization tools: a \textit{screen-based heatmap} showing density in screen space; an \textit{image-based heatmap} showing density after inverse-transform reconstruction; a \textit{trace overlay} connecting ordered gaze points; and a \textit{replay/simulation} tool that replays transformations and gaze in lockstep for qualitative verification.


\begin{figure}[ht]
  \centering
  \includegraphics[width=0.18\textwidth]{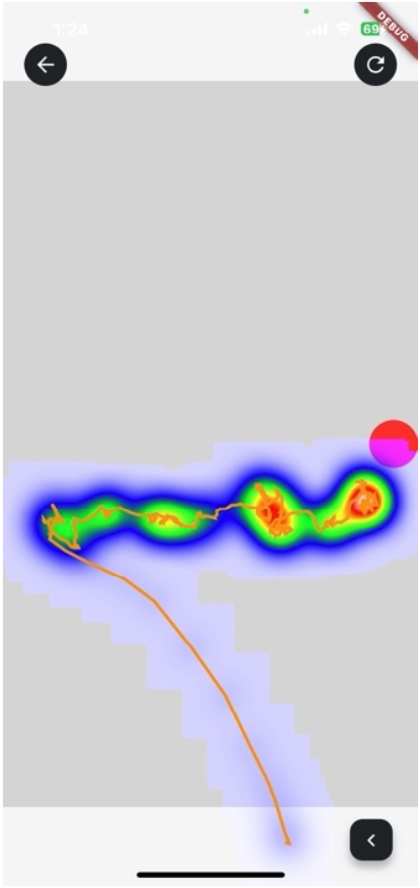}
  \hspace{0.02\textwidth}
  \includegraphics[width=0.18\textwidth]{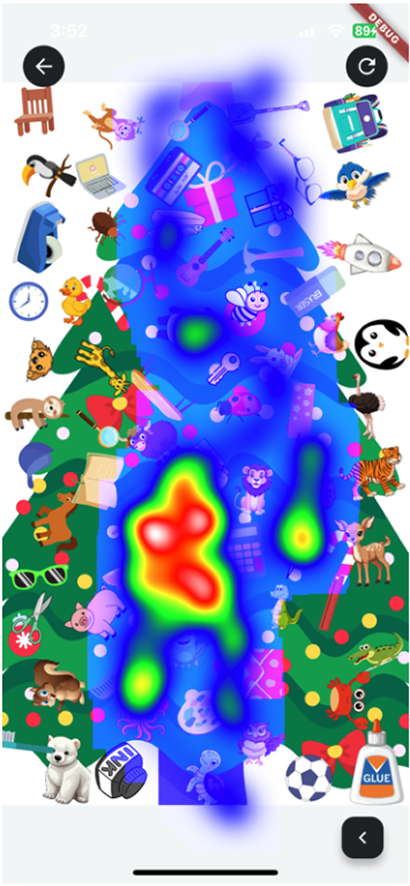}
  \caption{Screen-space vs.\ image-relative heatmaps. Under content manipulation, screen-coordinate gaze appears spatially dispersed, whereas inverse-transform reconstruction reveals gaze concentration on stable regions of the underlying image.}
  \Description{Two heatmaps shown side by side. The left panel is a screen-space heatmap where gaze density appears stretched or scattered due to content movement. The right panel is an image-relative heatmap over the original image, showing more coherent clusters aligned to specific content regions after reconstruction.}
  \label{fig:heatmaps}
\end{figure}

\section{Evaluation}

\subsection{Study Design and Procedure}
We conducted a formative technical validation with $N{=}3$ participants from the university population. The goal was not to generalize behaviour, but to verify the end-to-end pipeline (synchronization integrity, reconstruction plausibility, and replay usability) and to surface failure modes under realistic manipulation patterns. Sessions lasted approximately 30 minutes and included 12 tasks spanning guided manipulation, reading, and visual search, i.e. 12 tasks $\times$ 3.

The procedure followed a consistent flow: \textit{(i)} familiarisation with pinch/rotate gestures on a trial page, $\Rightarrow$ \textit{(ii)} five-point calibration using the SDK's built-in routine, $\Rightarrow$ \textit{(iii)} execution of 12 structured tasks spanning guided manipulation, reading, and visual search. To reduce uncontrolled posture variance during this initial validation, participants were instructed to hold the phone upright throughout the tasks. Recalibration was permitted whenever alignment visibly degraded, and these interventions were recorded as part of the session-level data-quality notes.

\subsection{Quantitative Log-Quality and Accuracy Slice}

To complement the qualitative replay/heatmap inspection, we report a small set of quantitative indicators from the guided tasks. These metrics are intended to characterise pipeline reliability and current operating boundaries, rather than to claim benchmark-level gaze accuracy. Specifically, we summarise (i) synchronisation success within the 50\,ms tolerance, (ii) discarded sample rate, (iii) recalibration frequency, and (iv) a simple guided-task error measure in image coordinates (distance from the instructed fixation target after reconstruction). For reference, we also report the corresponding screen-coordinate error to illustrate how screen-space analysis can be misleading under manipulation.

\begin{table}[ht]
\centering
\begin{tabular}{ll} 
\toprule
Metric                                             & Value    \\ 
\midrule
Matched samples within $\Delta{=}50$ms             & 96.5\%   \\
Discarded samples (no valid match)                 & 3.5\%    \\
Recalibrations per 30-min session (median [range]) & 3 [2–5]  \\
Guided-task image-coordinate error (median px)     & 92       \\
Guided-task screen-coordinate error (median px)    & 640      \\
\bottomrule
\end{tabular}
\caption{Quantitative summary of synchronization quality and guided-task reconstruction performance under iPhone 15 Pro Max usage.}
\Description{A two-column table with columns labeled Metric and Value. It reports five indicators from the formative validation: 96.5 percent of samples were matched within a 50 ms synchronization tolerance; 3.5 percent were discarded due to no valid match; the median number of recalibrations per 30-minute session was 3, with a range of 2 to 5; the median guided-task error after image-relative reconstruction was 92 pixels; and the corresponding screen-coordinate error was 640 pixels. The table highlights high synchronization success and a substantially lower error in image coordinates than in screen coordinates under content manipulation.}
\
\label{tab:quant_summary}
\end{table}

\subsection{Task Set and Validation Rationale}
The 12 tasks were designed to cover both \emph{ground-truth-like} and \emph{naturalistic} interaction: 
(Table~\ref{tab:tasks}).

\begin{table}[ht]
\centering
\small
\begin{tabular}{p{0.16\linewidth} p{0.48\linewidth} p{0.28\linewidth}}
\toprule
Category & Example tasks (from the 12-task set) & What it validates \\
\midrule
Guided manipulation & Move image along a dotted guide while fixating a target (e.g., right$\rightarrow$left; top$\rightarrow$bottom; diagonal; diamond); rotate within a boundary; semi-circular motion & Synchronisation; basic plausibility of reconstruction and replay \\
Reading & Read aloud / read vertically-oriented text while manipulating the image rather than rotating the device & Stability under moderate manipulation; edge-case behaviours \\
Visual search & Find small concealed objects in complex images (e.g., ``Crocodile''/comb) requiring extensive pan/zoom/rotation & Robustness under naturalistic, multi-step transformations \\
\bottomrule
\end{tabular}
\vspace{0.2em}
\caption{Task categories in the 12-task validation set and their role in evaluating the GazeSync pipeline.}
\Description{A three-column table summarizing the validation task design. The first column lists three categories: Guided manipulation, Reading, and Visual search. The second column provides representative tasks for each category, such as moving an image along dotted paths while fixating a target, reading text while manipulating the image, and searching for hidden objects in complex scenes. The third column explains the validation purpose of each category: guided tasks test synchronization and reconstruction plausibility, reading tasks test stability and edge cases under moderate manipulation, and visual search tasks test robustness under naturalistic multi-step transformations.}

\label{tab:tasks}
\end{table}

\begin{figure*}[ht]
  \centering
  \includegraphics[width=0.95\textwidth]{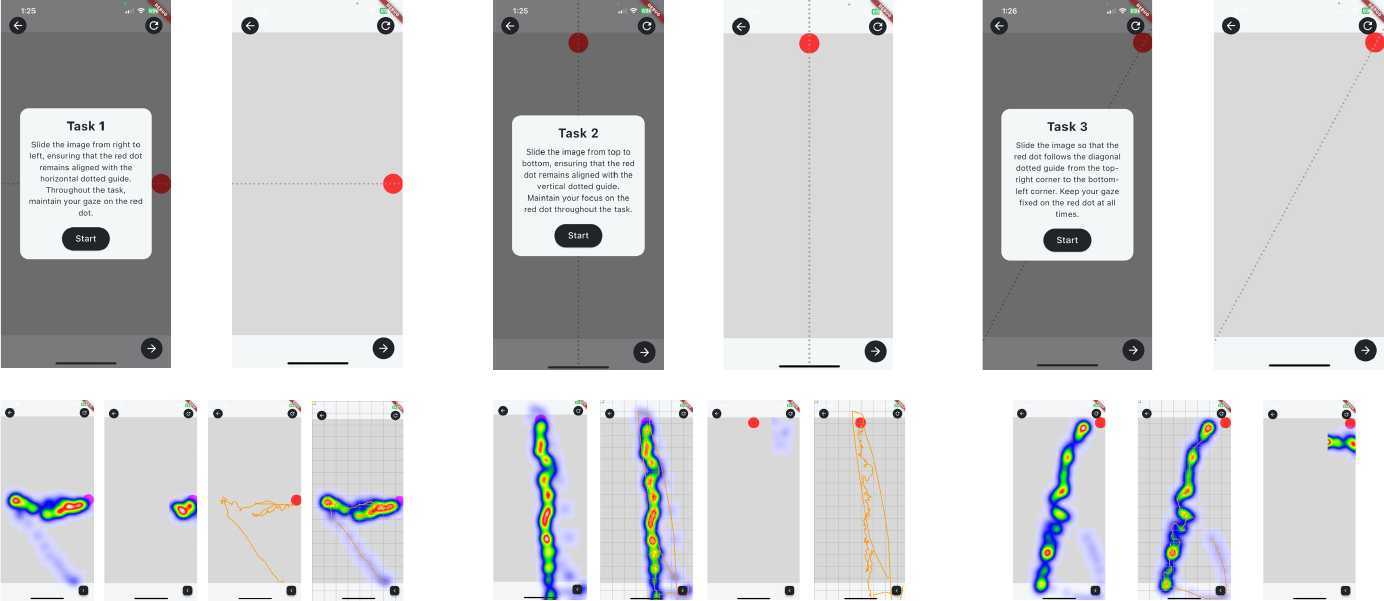}
  \caption{Guided manipulation tasks with line-based trajectories. Participants pan the image along constrained paths while maintaining fixation on a target marker. Representative task screens and corresponding analysis outputs (e.g., trace and heatmap overlays) are shown to validate synchronization and the plausibility of content-relative reconstruction.}
  \Description{A multi-panel composite figure showing examples of guided tasks based on line trajectories. The top row contains task screens with dotted guide paths and a red target marker. The bottom row contains corresponding analysis views, including gaze traces and heatmaps. The panels illustrate horizontal, vertical, and diagonal image movements while the participant keeps gaze on the target.}
  \label{fig:task1}
\end{figure*}

\begin{figure*}[ht]
  \centering
  \includegraphics[width=0.95\textwidth]{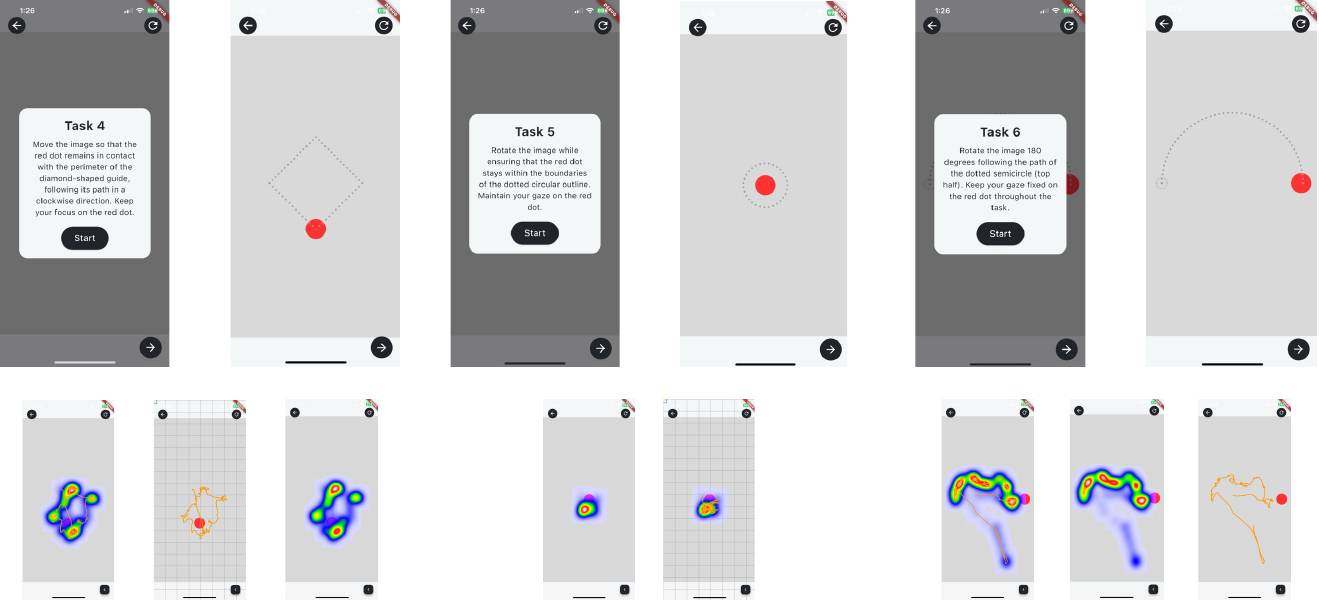}
  \caption{Guided manipulation tasks with shape-based trajectories. Motions such as diamond-shaped and arc-like paths increase interaction complexity and stress-test reconstruction under larger translations and curved movements. The accompanying analysis views support rapid inspection of gaze alignment with the intended target during manipulation.}
  \Description{A multi-panel composite figure showing guided tasks with shape-based movement patterns. The top row shows task screens with geometric paths, including a diamond path and a curved or arc-like path, each with a target marker. The bottom row shows analysis outputs such as heatmaps and traces. The figure demonstrates more complex motion patterns than the line-trajectory tasks.}
  \label{fig:task2}
\end{figure*}

\textbf{Guided manipulation tasks (Tasks 1--7).}
Participants follow explicit on-screen guides (e.g., dotted paths, bounded rotation, semi-circular motion) while maintaining fixation on a target marker (Figure~\ref{fig:task1} and Figure~\ref{fig:task2}). These tasks provide sanity checks for (a) synchronisation and (b) whether reconstruction yields plausible content-relative gaze.


\begin{figure*}[ht]
  \centering
  \includegraphics[width=0.95\textwidth]{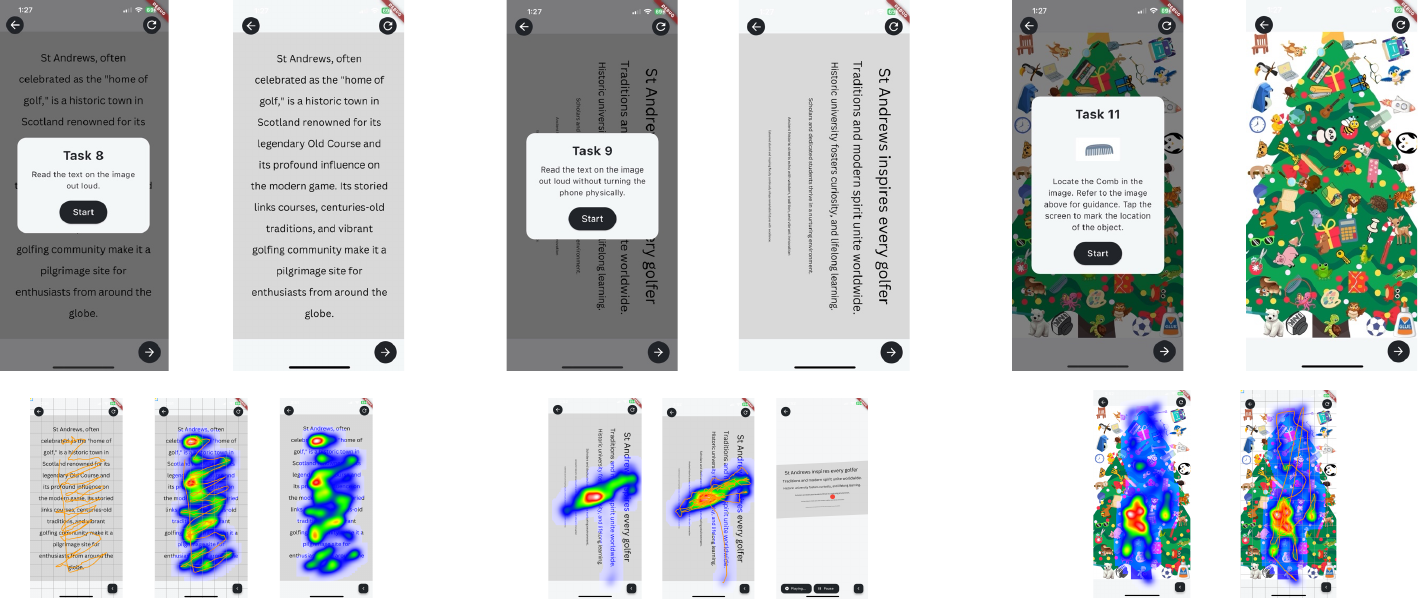}
  \caption{Naturalistic task examples (reading and visual search). Reading tasks (including rotated or vertical text) and visual search tasks elicit unscripted zoom--pan--rotate behaviour. Representative analysis outputs illustrate how replay and overlays help interpret attention patterns that are ambiguous in screen coordinates alone.}
  \Description{A multi-panel composite figure with naturalistic task examples. Panels include reading tasks with text in different orientations and a visual search task in a cluttered scene. Additional panels show analysis outputs, including gaze heatmaps and trace overlays. The figure highlights how user-driven manipulation during reading and search produces complex gaze patterns that benefit from content-relative analysis.}
  \label{fig:task3}
\end{figure*}

\textbf{Reading tasks (Tasks 8--9).}
Participants read text at different sizes/orientations, naturally inducing zoom/pan/rotation adjustments to support comfortable reading (Figure~\ref{fig:task3}, left).

\textbf{Visual search tasks (Tasks 10--12).}
Participants locate small targets hidden in complex scenes (``Where's Waldo''-style), inducing diverse, unscripted manipulation patterns (Figure~\ref{fig:task3}, right).



\section{Findings and Discussion}
GazeSync successfully logged synchronised gaze and transformation data for all participants across all tasks and enabled rapid qualitative validation via replay and heatmaps.

\textit{Ground-truth tasks verified synchronisation.} In guided movement tasks (Figures~\ref{fig:task1}--\ref{fig:task2}), the \textit{screen-based} heatmaps concentrated along expected screen regions, and replay showed gaze and manipulation evolving in a consistent timeline. In multiple instances, \textit{image-based} heatmaps correctly localised attention to the instructed target marker, supporting the core premise of content-relative reconstruction. \textit{Reading tasks produced interpretable gaze dynamics.} In the reading tasks (Figure~\ref{fig:task3}), replay commonly showed classical reading patterns (saccade--fixation sequences) while participants panned/zoomed text for comfort. In the vertical-text reading task, replay clarified that participants rotated the image (rather than the device), explaining otherwise confusing screen-space traces. 3) \textit{Integrated analysis accelerated debugging.} Being able to view traces, heatmaps, and replay immediately after a task made it easy to detect calibration failures and confirm whether logs aligned with expected behaviour.

\textbf{Key Limitations Observed.}
First, we observed \textit{calibration repetition and drift}. Consistent with single-camera mobile tracking literature~\cite{lei2025quantifying, lei2023dynamicread}, calibration drifted over the 30-minute sessions, requiring recalibration between task blocks. This suggests future systems should implement ``implicit calibration'' techniques, such as continue calibration~\cite{lei2025mac,lei2021eye}, calibration during interaction~\cite{cai2025gazeswipe, sugano2015appearance}, to correct drift on the fly. Second, we noted \textit{reconstruction failure under complex manipulations}. Image-relative heatmaps were fragile when participants performed heavy compound transformations (e.g., rapid rotations exceeding 45 degrees combined with repeated zooming), where small temporal offsets and rotation errors compounded. For example, in a visual search task, the in-app replay tool made it immediately apparent that during rapid, alternating pan-and-rotate gestures, the reconstructed gaze trailed behind the target by several frames before restabilising.

\textbf{Implications for Mobile Attention Research.}
GazeSync provides a reusable methodological toolchain for studying attention on interactive visual content beyond static screens. In particular, the unified data synchronization schema, inverse-transform reconstruction procedure, and in-app replay/heatmap workflow can be integrated into other mobile HCI studies that involve maps, photos, documents, or scans under touch manipulation. At the same time, our findings show that \emph{content-relative analysis depends not only on gaze estimation, but also on calibration maintenance and transform robustness}. This motivates future work on (i) continuous/adaptive calibration during interaction, (ii) transform-aware filtering and numerical stabilisation for inversion, and (iii) higher-level visualisations and metrics built on fixations/events rather than raw samples.

\section{Conclusion}
We introduced GazeSync, a mobile tool for \emph{content-relative} attention analysis during dynamic image interaction. By synchronising gaze samples with UI transformation states, GazeSync enables researchers to analyse what content users attended to under pan/ zoom/ rotate interactions, rather than relying only on gaze point on screen coordinate. This formative technical validation showed that the approach is practically useful for guided manipulation, reading, and visual-search tasks, while also making current limitations visible, particularly calibration drift and reconstruction fragility under heavy compound transforms. These findings position GazeSync as a reusable methodological foundation for mobile HCI studies, and they motivate the next development steps: continual/implicit calibration, transform-aware stabilisation, and fixation-level analysis for cleaner and more robust interpretation.

\begin{acks}
We thank all participants for their time and contributions. We also thank the CHI reviewers for their constructive feedback, which helped improve the clarity and presentation of this work. Finally, we are grateful to our colleagues for helpful discussions and support throughout the project.
\end{acks}

\bibliographystyle{ACM-Reference-Format}
\bibliography{reference}


\clearpage
\appendix

\end{document}